\documentclass[twocolumn,showpacs,preprintnumbers,amsmath,amssymb]{revtex4} 
\usepackage{graphicx}% Include figure files
\usepackage{dcolumn}% Align table columns on decimal point
\usepackage{bm}% bold math
\begin{document}

\preprint{CUPhys/1/2007}

\title{Dynamics of unvisited sites in presence of mutually repulsive random walkers}

\author{Pratap Kumar Das}
\author{Subinay Dasgupta}
\author{Parongama Sen}%
\affiliation{%
Department of Physics, University of Calcutta,92 Acharya Prafulla Chandra Road,
Calcutta 700009, India.\\
}%

\date{\today}% It is always \today, today,

\begin{abstract}

We have considered the persistence of unvisited sites of a lattice, 
i.e., the probability $S(t)$ that
a site remains unvisited till time $t$  in presence of mutually repulsive 
random walkers.
The dynamics of this system has direct correspondence to that 
of the domain walls in a
certain  system of Ising spins where the number of domain walls 
become fixed following a zero termperature quench.
Here we get the result that $S(t) \propto \exp(-\alpha t^{\beta})$
where $\beta$ is close to 0.5 and $\alpha$ a function of the
density of the walkers $\rho$.
The number of persistent sites in presence of independent walkers of density 
$\rho^\prime$  is known to be $S^\prime (t) =
\exp(-2 \sqrt{\frac{2}{\pi}} \rho^\prime t^{1/2})$.
We show that a mapping of the interacting walkers' problem to the 
independent walkers' problem is possible  with $\rho^\prime = \rho/(1-\rho)$ 
provided $\rho^\prime, \rho$ are small.
We also discuss some other intricate results obtained in the 
interacting walkers' case.

\end{abstract}

\maketitle

\section{Introduction: The original spin problem}

Dynamical evolution of a spin system following a quench to zero
temperature from a disordered state may lead to a non-equilibrium state,
e.g., as in the one dimensional ANNNI (Axial Next Nearest Neighbor Ising)
 model \cite{Selke}
 which has the Hamiltonian
\begin{equation}
H = -\Sigma S_iS_{i+1} + \kappa\Sigma S_iS_{i+2},
\end{equation}
where  $S_i = \pm 1$ is the spin at the $i$th site and $\kappa >0$ is the ratio
of the second neighbour and first neighbour interactions.  
For $\kappa < 1$, the quench  does not lead to the 
equilibrium configuration which is ferromagnetic
for $\kappa < 0.5$ and antiphase for $\kappa > 0.5$. 
Starting from a random state, there is a short initial time
 during which the domains of size one
 die  and this eventually results in a configuration with fixed number of
domain walls. In this  state,
the  domain walls  become
 ``fluid'' in the sense that they can move indefinitely (keeping
the energy of the system same) \cite{Redner,Sen1,Sen2} but cannot cross each other.
In such a system the persistence dynamics shows that 
the number of persistent spins is neither a power law nor 
exponential but rather follows a stretched exponential 
decay, 
\begin{equation}
P(t) \sim \exp(-\alpha t^{\beta}),
\end{equation}
 with $\alpha \approx 1$ and
$\beta=0.45$ \cite{Sen1}.

The above dynamical scenario can easily be represented by an
equivalent system of mutually avoiding random walkers and the fraction of 
persistent
spins will then be given by the fraction of unvisited sites $S(t)$ till time $t$
in the system \cite{Fisher,Huse,Derrida,Molina,Yuste,Dfisher,Grassb,Derrida2,Forrester,Kratten,Katori,Bray,Bhatt}. Representation of the spin dynamics by random walk of domain walls 
is well-known; in the common examples like the Ising or Potts model, the 
random walk is accompanied by annihilation of two domain walls if they meet
\cite{Derrida2,Forrester,Bhatt}.

In the original ANNNI model problem, when there are $M$ domains 
with each separated by at least two lattice spacings, the probability distribution 
of the number of domain walls $M$ within a size $L$ is  
\begin{equation}
      {P(L,M)} =  {{L-M} \choose {M}} /  {\sum _ {M}^{L/2} {{L-M} \choose{M}}}.
\end{equation}
This equation is easy to derive once it is realised that the 
problem is   identical  to the Bose statistics of  distributing 
$L$ particles in $M$
boxes with the number of particles in each box greater than or equal to 2.
It has been observed that the ratio $\rho_0 = M/L$ has a mean  value quite 
close to the most probable value $ \rho_0 = 0.2764$ \cite{Sen1}. 
Distribution of the value of $ \rho_0$ can be calculated numerically 
which shows that its fluctuation decreases with the system size 
(typically $\Delta \rho_0$ decreases from 0.0645 for $L=20$ to 0.0225 for
 $L=175$ in a power law manner, $\Delta\rho_0\sim N^{-0.5}$). This indicates 
that the ANNNI dynamics
reflects the behavior of $S(t)$ for a specific value of the density of walkers
in the equivalent random wall picture. It is therefore a meaningful exercise
to find out $S(t)$ for general density $\rho=N/L$ where 
$N$ is the number of mutually exclusive walkers and
$L$ the chain length. In this paper we have considered $0 < \rho < 1$ and
 calculated numerically $S(t)$. 
$S(t)$ shows a  stretched exponential behaviour with an exponent close 
to 1/2.

This problem, as we have shown, can be mapped to that of 
the   independent random walkers with a subtle difference
and also for small $\rho$.
 The
latter problem has been exactly solved \cite{bray} and we compare its 
results with that of the numerical simulation of mutually repulsive walkers
to find a good agreement at small values of $\rho$. 

Our results also show a   difference when persistence is 
calculated in terms of a   spin system and the system of pure brownian 
walkers which we have  discussed in the paper.
Some other dynamical properties have been investigated in this context.

\section{Persistence of unvisited sites in presence of  mutually exclusive walkers}

To keep correspondence with 
the original spin dynamics of the ANNNI model, any of the 
site in the system may be selected for updating. It may or may not contain a 
random walker.   If it does, then the step taken by this random walker 
is according to the following three situations:\\
%\newpage
1. If the walker is flanked by two random walkers on both
sides; it cannot make any movement and stays there - this corresponds to
a locked domain wall (Fig 1a).  \\ 
2. There are  no   neighbouring random walker, then 
the random walker  remains at its position with probability $1/2$ - this corresponds 
to the probability that a spin does not flip even at the domain boundary.
It can also move to either left or right with equal probability (Fig 1b).\\
3.  There is only one neighbouring  walker, say, to  the left (right), 
then it does not  move or   
 moves  to the right (left) (Fig1 (c) and (d)) with equal probability.\\
Let us represent this dynamical rule by $D_{ANN}$, a dynamics which corresponds to the ANNNI model dynamics.

As the  walls are reflecting,
if the  random walker happens to hit the  wall, it can only step
in-wards or stay there. 
When $L$ sites are hit, one Monte Carlo step is said to be
completed. 

\begin{center}
\begin{figure}
\noindent \includegraphics[clip,width= 4cm, angle=0]{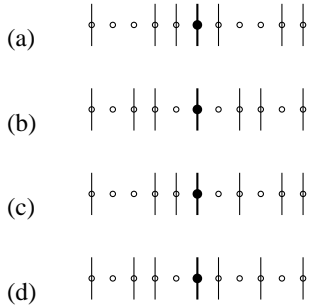}
\caption{Possible movement of the tagged walker (highlighted): $(a)$ cannot move. $(b)$
either does not move or can move to the left/right. $(c)$ either does not
move or can move to the right. $(d)$ either does not move or can move to the left.}
\end{figure}
\end{center}

It may be noted that in the original ANNNI model, domain walls need to 
maintain a least distance of two lattice spacings. We have relaxed this condition
to one lattice spacing 
which is equivalent to having on-site repulsion of
random walkers. 
Effectively, this means that
the original ratio $\rho_0 =M/L$ in ANNNI corresponds to $\rho=N/L=2\rho_0$ in the
present case.

The starting position of the random walkers may be assumed  to be either
already visited or not yet visited. The calculation of persistence
will depend on it. In what we call the spin picture (SP), they are unvisited and
in the random walker picture (RWP) they are assumed to be visited already.
This brings in a subtle difference in the two problems. 
We have discussed the two problems separately in the following two subsections.

\subsection{The spin picture}

First, we calculate the  survival probability $S(t)$ defined as the probability 
that a site has not been visited by any of the random walkers till time $t$.  
(This corresponds to $P(t)$, the persistence probability of the
original ANNNI model with $\rho \simeq 0.54$.)

\medskip
In Fig. 2, $S(t)$ is plotted against time $t$ for different densities $\rho$
with a fixed lattice size $L=10000$. In each case, $S(t)$ decays with
time following a stretched exponential behavior $\exp(-\alpha t^{\beta})$
where $\beta = 0.50 \pm 0.02$. This value of $\beta$ compares well
with $0.45$  obtained for $P(t)$ in the ANNNI model \cite{Sen1}.
Increasing the system size does not affect the result, only the 
variation of $S(t)$ can be observed over a longer duration of time.

\begin{center}
\begin{figure}[b]
\noindent \includegraphics[clip,width= 6cm, angle=270]{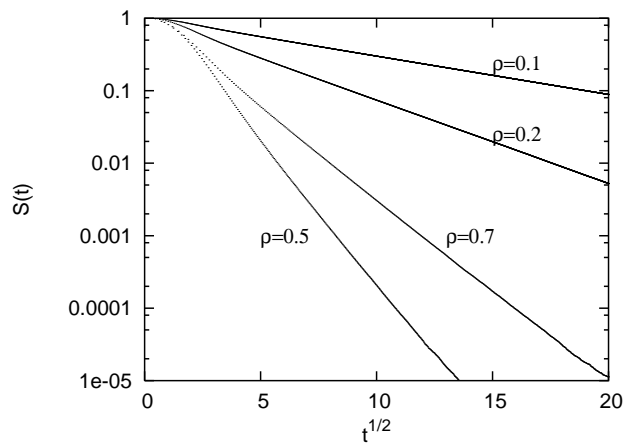}
\caption{$(a)$ Survival probability $S(t)$  as a function of time $t$ 
for different densities of walkers on a lattice of size 10000. 
Each curve follows a stretched
exponential of the form  $\exp(-\alpha t^{0.50})$.}
\end{figure}

\end{center}

Interestingly,
the value of $ \alpha $ increases with the the increase of the density $\rho$
 till the value of $\rho \approx 0.55 $.
Beyond this value, $ \alpha $ decreases gradually as the density 
increases 
roughly as $(1-\rho)$ (Fig. 3).
The behaviour of $\alpha$ as a function of $\rho$ will be discussed in greater 
detail in the next section.

The qualitative behaviour of $\alpha$ as a function of $\rho$ is not difficult to explain. For small values of $\rho$,
the probability that a domain wall is hit is small and therefore one gets
a slow decay  of $S(t)$ with time. On the other hand, for large values of $\rho$,
most of the domain walls will be `locked' such that again the variation of $S(t)$ will be
slow.  

For very small $\rho$, we have checked  that
\begin{equation} 
\label{twice}
 \alpha(1-\rho) = \sqrt{2} \alpha(\rho) 
\end{equation}
holds good
 to a high degree of
accuracy. Effectively this means that
the relaxation rate at $\rho \to 1$ is twice as much as that at $\rho \to 0$.
This can be explained in the following manner: 
At  $\rho \to 1$, an empty site occurs with a probability $(1-\rho)$
but
has both neighbours occupied 
 with a very high probability, almost one.  At $\rho \to 0$, an empty site
 having  an occupied   neighbour is very rare (probability proportional to $\rho)$.
In the former, in one iteration, the empty site can be visited by either of its neighbouring walkers while in the latter,
visit to the empty site is possible by one walker only, making the time scale double.

\begin{center}
\begin{figure}
\includegraphics[width=6cm,angle=270]{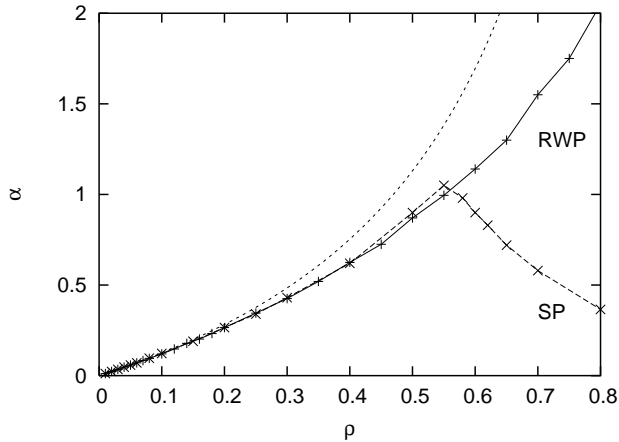}
\caption{ Slope of the stretched exponential curve $\alpha$ is plotted
against the density $\rho$ for both the SP (spin picture) and RWP (random walkers picture) for the  dynamical rule $D_{ANN}$  (see section II).
Both the curves follow a behaviour $1.13\rho/(1-\rho)$ for small $\rho$, 
shown by the dashed line.}
\end{figure}
\end{center}

One can now compare the value of $\alpha$ obtained in the present study
with that of the ANNNI model where $\rho \simeq 0.27$. In \cite{Sen1},
the value of $\alpha$ was found to be equal to 1.06.
  For $\rho=2\rho_0 \simeq 0.54$, we find 
that $\alpha$ is very close to the value 1 (Fig. 3) showing again 
a good agreement
with the ANNNI model dynamics.

\subsection{The Random walker picture}

In the RWP  everything remains same but the initial sites occupied by the random walkers 
are assumed to be already visited. Now we find that $S(t)$ again has a stretched exponential
behaviour : $S(t) \sim \exp(-\alpha_{RW} t^{\beta_{RW}})$
with  $ {\beta_{RW}} = 0.5 \pm 0.01$.
Now $\alpha_{RW}$ does not show any non-monotonic behaviour but appears to diverge as
$\rho \to 1$. This is again understandable, in the present picture, 
when $\rho$ is close to one,
most of the sites are already non-persistent to begin with and $S(t)$ decays very fast making $\alpha_{RW} \to \infty$.

The exponents $\beta_{RW} $ and $\beta$ are apparently equal in the 
two pictures.
In Fig. 3 the behaviour of $\alpha_{RW}$ against $\rho$ is also shown. 
It is to be noted that upto $\rho\approx 0.5$, $\alpha $ and $\alpha_{RW}$
 are equal and behave differently beyond this point. For the SP, the possibility of 
the domains getting locked increase as $\rho$ increases and this happens with a higher probability
beyond $\rho = 0.5$.

\section{Mapping to a system of non-interacting walkers}

In the last section we obtained the result that the fraction of unvisited 
sites $S(t)$ in presence of mutually repulsive walkers has a stretched
exponential decay with exponent 1/2  in both the SP and RWP.
This behaviour turns out to be  exactly the same as that of 
$S^\prime(t)$,  the number of unvisited sites  in the presence of 
independent or non-interacting walkers. In the latter system it has been shown \cite{bray} that
 when $\rho^\prime$ is the density of independent walkers,  
\begin{equation}
S^\prime (t) \sim \exp(-\alpha^\prime \rho^\prime \sqrt{t}),
\end{equation}
with $\alpha^\prime= 2\sqrt{\frac{2}{\pi}}$. 

In this section we show that the interacting system can be mapped to the independent walkers' system
with the transformation $\rho^\prime = \rho/(1-\rho)$. 
To show this, let us consider a configuration $C$ of $N$ interacting walkers 
of density $\rho$ which follow a  dynamics represented by $D$.
For the present discussion,
we make the dynamics $D$ simpler than $D_{ANN}$: 
the walker will always execute a movement when at least one of
the neighbouring site is vacant - if both are vacant, probability to move 
either to the  left or to the right
is 1/2. In case only one neighbouring site  is occupied, the random walker 
will move to the empty neighbouring site.
(One obtains the same behaviour of $S(t)$ with this rule, including the relation (\ref{twice}),
only
the numerical value of $\alpha$ increases  by a factor of $\sqrt{2}$ compared to $D_{ANN}$ where the time scale is simply double compared to $D$.)

For the independent walkers, let us consider a configuration $C_0$ of 
$N$ walkers of density $\rho^\prime$, 
who do not ``see'' each other. The dynamics $D_0$ here is simply that each walker will
move to left or right with equal probability.
The world lines in the 1+1 dimension of the walkers are shown in Fig 4a and 4b.

Now let us create a mapping of the original configuration $C$ to $C^\prime$
given by
\begin{equation}
x^{\prime}_k(t) = x_k(t) -k,
\end{equation}
where $x_k(t)$ is the position of the $k$th walker ($k=1,2,...M$ from the left) at time $t$ \cite{villain}.
Effectively this mapping implies  that one spacing between consecutive walkers is being removed.
This would remove the constraint in $C$ that two walkers have hard core repulsion and 
each world line of $C^\prime$ therefore also occurs in $C_0$. Although all world
lines of $C^\prime$ and $C_0$ have one to one correspondence, in $C^\prime$ one has the
constraint that $x_1 < x_2 < x_3.....<x_N$ while in $C_0$ there is no such constraint. Therefore a particular configuration may occur with  different
weight factors in $C_0$ and $C^\prime$ (Figures 4b,c, and d).

\begin{center}
\begin{figure}
\includegraphics[width=7cm]{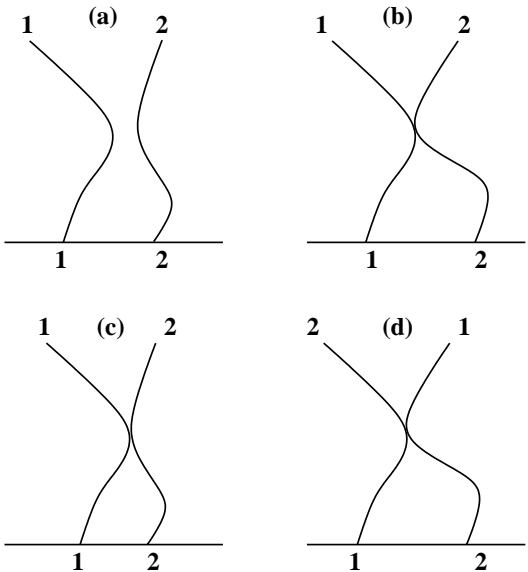}
\caption{Typical examples of (a)  mutually repulsive walkers ($C$) and (b) Independent walkers 
($C_0$). (c) The mapped configuration $C^\prime$ following  eq. 6.  
In (d), another realisation  for the independent walkers is possible by
switching the order of the walkers at the later time. This is not allowed in $C^\prime$.}
\end{figure}
\end{center}

Using the above mapping, the effective chain length in $C^\prime$ is $L-N$  and not $L$. Thus 
the density $\rho$ in $C$ is related to the density $\rho\prime$ in $C^\prime$ 
by the equation
\begin{equation}
\rho^\prime = \frac{\rho}{1-\rho}.
\end{equation}

Ignoring the weight factor, the mapping is effective in 
showing the correspondence between
the interacting and independent walkers' picture. Exact correspondence will imply that   
 $\alpha$ or $\alpha_{RW}$ (with $D$) would be equal to  $\alpha^\prime (\frac{\rho}{1-\rho})$, where $\alpha^\prime =  2 \sqrt{(2/\pi)}$, when dynamics $D$ is
used. This can happen if    
the dynamical rule $D$ applied to $C$ leads to states 
which when mapped to $C^\prime$ will correspond  exactly to the states 
obtained by applying $D_0$ on $C_0$ (with the same weightage).   
We have verified that this is true  for configurations  in which either a walker is ``alone''
(both neighbours are empty)  or has at most one walker in a neighbouring site.
However, when three walkers occupy consecutive sites (a ``three'' state), 
the dynamics $D$ 
gives rise to states which cannot be obtained from $C_0$ applying $D_0$ on it.
Since the probability of having ``three'' states increases with $\rho$, we expect that the results for independent and
interacting walkers will differ at higher $\rho$. 
For $D_{ANN}$, it is expected that  $\alpha$ and 
$\alpha_{RW}$ values would  be   equal to
$\frac{\alpha^\prime}{\sqrt{2}}\rho/(1-\rho) = \frac{1.13\rho}{1-\rho}$ upto small $\rho$ which is exactly 
what we observe (Fig. 3) (time scales for $D_{ANN}$ being simply twice that of $D$).
%The numerical results confirm this, as we notice that $\alpha$ or ($\alpha_{RW}$) can be fitted by 
%$ \frac{1.13\rho}{/1-\rho}$ only upto small values of $\rho$ (Fig. 3).

Obviously a ``three'' state cannot be avoided if $\rho > 2/3$ and this
gives an upper bound where the disagreement will occur. 
In reality, such states occur at values of $\rho$ much below than this,
even  at about $\rho=0.2$. We have
verified that, if the three-states are forcibly ruled out in the simulation,
the correspondence between the independent and interacting walkers 
remain valid upto  $\rho\approx 0.4$.
 
We would like to comment in this section that 
while for the interacting walkers' system, $\alpha$ behaves differently
in the SP and RWP, no such difference exists for the independent
walkers' case. This is because  there is no restriction
on the movement of the walkers here even as the density becomes high.

A subtle point relevant to the mapping
needs to be mentioned here. At small $\rho$ the results for  
$C$ and $C_0$ are equivalent indicating that the dynamical evolution of the
walkers can be mapped to each other. It may still remain a question whether 
the persistence probability 
of $C$ can be mapped to that of $C^\prime$.
The question arises as 
the  $N$ sites removed from the original system may either be 
persistent or non-persistent.  
On an average, however, the persistence of the two systems $C$ and $C^\prime$ 
will
be same.
This is because the average number of 
persistent sites removed is $P(t)N$. Thus in the mapped system, persistence
probabilty is again $(P(t)L -P(t)N)/(L-N) = P(t)$. 
This justifies the correspondence of persistence in $C$ and $C^\prime$ and hence $C_0$.
The issue of equivalence of persistence requires special 
mention  as persistence 
is not related to other dynamical
behaviour of a system in general.   

\section {Non-monotonicity of $\alpha$ and a few relevant comments}

The result for persistence probability in the 
spin picture and random walker picture differ in the interacting walkers' case as in the SP there is a non-monotonicity
in $\alpha$.
This non-monotonic behaviour is clearly due to two features (a) presence of interacting
walkers and (b) the dynamic quantity under consideration  being persistence.

Point (a) is already discussed  in the last  section.
Regarding point (b), it must be noted that the non-monotonicity 
appears when we assume that the initially occupied points are not visited, a fact which is relevant
to persistence dynamics only. In this section we have discussed a few other dynamical 
phenomena in presence of  interacting walkers. However, we find that none of these are
accompanied by any non-monotonic behaviour of the relevant quantities appearing in them.

Two dynamic quantities $\sigma_1$ and $\sigma_2$ representing 
fluctuations  can be defined 
in the following way:
we tag a random walker and 
calculate the fluctuation
of its position $x(t)$ at time $t$ with respect to its 
initial position $x(0)$ and 
study its behavior with
time (Fig. 5).

\begin{center}
\begin{figure}
\includegraphics[width=5cm,angle=0]{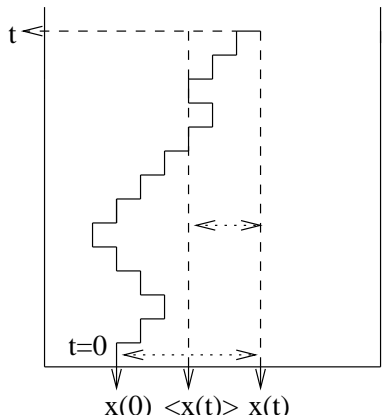}
\caption{ The movement of a tagged domain wall along time (vertical axis):
$x(0)$ is its initial position, $x(t)$ its position at time $t$ and
$\langle x(t) \rangle$, the mean position averaged up to time $t$.}
\end{figure}
\end{center}
Precisely, $\sigma_1$ is defined as 
\begin{equation}
\sigma_1(t) = \sqrt {\langle (x(t) -  x(0))^2 \rangle} . 
\end{equation}
In the second measure, we notice that the path of a walker can be viewed as
an  interface (with no overhangs) in 1 + 1 dimensions (Fig 5). 
One can then measure the interface width at any
time given by
\begin{equation}
\sigma_2(t) = \sqrt{(\langle x^2(t) \rangle - \langle x(t) \rangle^2)}. 
\end{equation}
where $\langle x(t) \rangle$ is the mean value of the position $x$ at time t.
It is known that for a single walker (i.e., $\rho=0$ in the thermodynamic
limit) $\sigma_1(t) = At^{\theta}$ with $\theta=0.5$. Here,
 we find  $\sigma_1(t) = t^{\theta}$ with $\theta\simeq 0.25$ at long times for all values of $\rho$.
This is in agreement with \cite{arratia} where the result $\theta=0.25$ has been
derived exactly. In the present system, $\rho$ has a finite value and for the smallest
value of $\rho$ shown in Fig. 6, a crossover effect is noted, i.e.,
the behavior at earlier time appears to be consistent with
$t^{0.5}$.
This is because at small $\rho$, the walker continues as a free walker
for a considerable period of time and exhibits the corresponding behaviour.

\begin{center}
\begin{figure}
\noindent \includegraphics[clip,width= 6cm, angle=270]{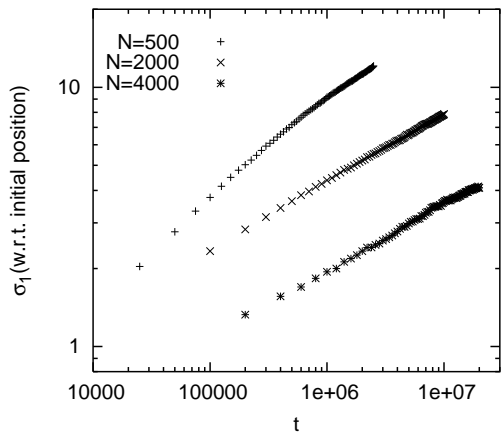}
\caption{Fluctuation $\sigma_1$ of the position of a walker w.r.t. its initial
position as a function of time $t$ for different no. of walkers
on a lattice of size 5000. The best fit curves have a slope $\simeq 0.25$ for the
higher densities.}
\end{figure}
\end{center}
The behavior of $\sigma_1(t)$ with time $t$ has been studied for 
values of $\rho$ even smaller than
0.1 in smaller lattices and it appears that for any finite $\rho$,
however small, $\theta\simeq 0.25$ is valid at longer times always.
We conclude that there is a transition
point at $\rho=0$ for any $\rho\ne 0$, the random walker exponent is $\simeq 0.25$
while for $\rho= 0$, it is 0.5.
The results for $\sigma_2$ are consistent with the above observations. 
Both $\sigma_1$, $\sigma_2 = At^{\theta}$ with $\theta=0.25$ independent 
of $\rho$ (for $\rho\ne0$), while $A$ depends on $\rho$. In Fig. 7, we plot $A(\rho)$
against $\rho$ for both $\sigma_1$ and $\sigma_2$. $A(\rho)$ decereases 
monotonically with $\rho$ and follows a rough exponential decrease as 
$A(\rho)\simeq \exp(-2\rho)$ except for values very close to 1, where one can
expect anomalous behavior.

We investigate the behavior of another quantity $D(t)$, which we define as 
the fluctuation of the distance $d(t)$ between two neighbouring walkers 
at time $t$ with respect to its initial value $d(0)$. Precisely,
\begin{equation}
D(t)=\sqrt{\langle(d(t)-d(0))^2\rangle}.
\end{equation}
$D(t)$ shows an initial increase  with $t$ and reaches a time independent 
equilibrium value $D_{sat}$ at larger times. This equilibrium value $D_{sat}$
(calculated from the mean value of the last 100 Monte Carlo steps), when 
plotted   against $\rho$, again shows a monotonic decay with $\rho$ (Fig. 7).
Hence, in contrast to the factor $\alpha$ appearing in the persistence
dynamics, we do not find non-monotonic behavior in  the 
factors appearing in the other dynamical features.

\begin{center}
\begin{figure}
\noindent \includegraphics[clip,width= 7cm, angle=270]{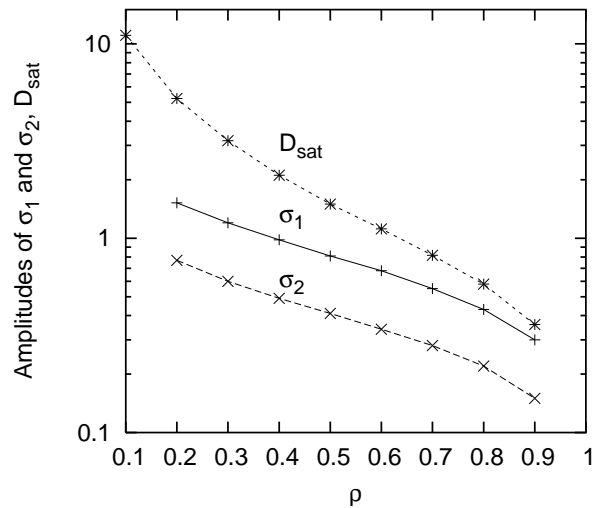}
\caption{Amplitudes ($A$) of  $\sigma_1$,$\sigma_2$($=At^{\theta}$) and
$D_{sat}$ as function of $\rho$. The dashed lines are guides to the eye.}
\end{figure}
\end{center}

\section{Summary and conclusions}

In summary, we have considered the dynamics of $N$ mutually avoiding
random walkers 
on a one dimensional chain of length $L$ which simulates the quenching dynamics in the ANNNI model for a particular value of $\rho\simeq 0.54$. 
For this value of $\rho$, we verify that the survival 
probability $S(t)$ which corresponds to the persistence probability 
in the ANNNI model follows a stretched exponential behavior
consistent with the ANNNI model dynamics. We also observe
very  good quantitative agreement for the 
exponent $\beta$ and slope $\alpha$. 
On generalising the value of $\rho$, we find that the behavior $S(t) \sim
\exp(-\alpha t^{\beta})$ is valid for all $\rho$ with $\beta$ showing a universal
value of $0.50\pm .02$. 

Observing that the time dependence  in $S(t)$ is identical to
that in $S^\prime(t)$ 
(the corresponding quantity in presence of 
non-interacting walkers) 
for all $\rho$, 
we have shown that a mapping between the two indeed 
exist which remains exact for small $\rho$ as far as the behaviour of $\alpha$ is concerned. 

We have considered two different pictures while computing the 
persistence probability; in the spin  (random walker) picture the 
sites initially
occupied by the walkers are assumed to be not visited (visited) and
the behaviour of $\alpha$ as a function of $\rho$ 
is sensitive to this difference.
In the SP, it has a non-monotonic behaviour. Such
non-monotonic behaviour emerges in the case of  
interacting walkers  only. However, when other dynamical phenomena 
in presence of interacting walkers are
studied, no such non-monotonic behaviour is found.
Thus we find that the persistence dynamics in a system with a finite
density of mutually avoiding random walkers calculated  in terms 
of the fraction of sites unvisited  till time $t$, shows a unique
behavior compared to other dynamical quantities. This again supports
the fact that persistence is a phenomenon which cannot be directly
connected to  other dynamical features of a system.

Acknowledgments: We are grateful to R. Rajesh for very helpful 
comments. We also acknowledge discussions with  S. N. Majumdar and P. Ray.
 P.K. Das acknowledges support from CSIR grant no.
9/28(608)/2003-EMR-I. P. Sen acknowledges support from CSIR grant no.
03(1029)/05-EMR-II. Financial support from DST FIST for computational
work is also acknowledged. 

%\newpage
\vskip 1cm
Email:  pratapkdas@gmail.com, sdphy@caluniv.ac.in, psphy@caluniv.ac.in

%\end{multicols}

\end{document}